\documentclass[a4paper,fleqn,usenatbib,useAMS]{mnras}

\usepackage{graphicx}	
\usepackage{amsmath}	
\usepackage{mathtools}
\usepackage[usenames, dvipsnames]{color}

\newcommand{\be}{\begin{equation}}
\newcommand{\ee}{\end{equation}}
\newcommand{\ba}{\begin{eqnarray}}
\newcommand{\ea}{\end{eqnarray}}
\newcommand{\eps}{\varepsilon}
\newcommand{\fracb}[2]{\left(\frac{#1}{#2}\right)}

\usepackage[T1]{fontenc}
\usepackage{ae,aecompl}

\usepackage{txfonts}

\title[Effect of Pair Cascades on Spectral Cutoff in GRBs]
{The Effect of Pair Cascades on the High-Energy Spectral Cutoff in Gamma-Ray Bursts}

\author[Gill \& Granot 2017]{Ramandeep Gill$^{1,2}$\thanks{Contact e-mail:
\href{mailto:rsgill.rg@gmail.com}{rsgill.rg@gmail.com}}
and Jonathan Granot,$^{1}$\thanks{Contact e-mail:
\href{mailto:granot@openu.ac.il}{granot@openu.ac.il}}
\\
$^{1}$Department of Natural Sciences, The Open University of Israel, 
1 University Road, PO Box 808, Raanana 4353701, Israel \\
$^{2}$Physics Department, Ben-Gurion University, P.O.B. 653, 
Beer-Sheva 84105, Israel}

\date{Last updated; in original form}

\pubyear{2017}

\begin{document}
\label{firstpage}
\pagerange{\pageref{firstpage}--\pageref{lastpage}}
\maketitle

\begin{abstract}
The highly luminous and variable prompt emission in Gamma-Ray Bursts (GRBs) arises in an ultra-relativistic outflow. The 
exact underlying radiative mechanism shaping its non-thermal spectrum is still uncertain, making it hard to 
determine the outflow's bulk Lorentz factor $\Gamma$. GRBs with spectral cutoff due to pair production ($\gamma\gamma\to e^+e^-$) at energies $E_c\gtrsim10\;$MeV 
are extremely useful for inferring $\Gamma$. 
We find that when the emission region has a high enough compactness, then as it becomes optically thick to scattering, Compton downscattering by non-relativistic $e^\pm$-pairs can shift the spectral cutoff energy well below the self-annihilation threshold, $E_{\rm sa}=\Gamma m_ec^2/(1+z)$. We treat this effect numerically 
and show that $\Gamma$ obtained assuming $E_c=E_{\rm sa}$ can under-predict its true value by as much as an order of magnitude.
\end{abstract}


\section{Introduction}


The GRB prompt emission is typically highly variable, consisting of multiple spikes
spanning a wide range of widths,
$\Delta T \sim 10^{-3}-1\;$s \citep[e.g.][]{FM95}. In the GRB central engine frame (CEF; cosmological rest frame of source) 
at redshift $z$, 
the variability time is $T_v = \Delta T/(1+z)$. For a Newtonian source light travel effects imply a source 
size $R\lesssim cT_v$. Since GRBs are extremely luminous sources, with typical energy fluxes 
$F\sim10^{-6}~{\rm erg~cm}^{-2}{\rm s}^{-1}$, and luminosity distances $d_L\sim10^{28}~$cm, a typical photon 
near the $\nu F_\nu$ peak with energy $E\sim E_{\rm pk}\sim m_ec^2$ would see a huge optical depth 
$\tau_{\gamma\gamma}(E)\sim\sigma_Tf_{\gamma\gamma}(E)n_\gamma R\sim10^{13}$ to pair production, 
$\gamma\gamma\to e^+e^-$ \citep{Piran99}, where $\sigma_T$ is the Thomson cross-section, $n_\gamma$ is the 
photon number density, and $f_{\gamma\gamma}(E)$ is the fraction of photons that can pair produce with the 
test photon of energy $E$.  This would imply a huge compactness $\ell\equiv\sigma_TU_\gamma R/m_ec^2$ 
(Thomson optical depth of pairs if all photons pair produce), where $U_\gamma$ is the radiation field energy density,
which would result in a nearly black-body spectrum, in stark contrast with the observed GRB non-thermal spectrum.

The solution to this so-called ``compactness-problem,'' is that the emission region must be moving 
towards us  ultra-relativistically with $\Gamma\gtrsim 10^2$ \citep[][]{Ruderman1975,Goodman1986,Paczynski1986,RM92}. 
\textcolor{black}{This implies: (i) Doppler factor: a blueshift such that the observed energy of photons 
$E = \Gamma E'/(1+z)$ (primed quantities are measured in the outflow's comoving rest frame) 
is higher by a factor of $\sim\Gamma$ than that in the comoving frame 
and (ii) the emission radius can be larger by a factor of $\sim\Gamma^2$, and assume a value of up to\footnote{In this work we adopt the convention 
$Q_x = Q/10^x$ (c.g.s. units).} \citep[see][for a review]{KZ15}
\begin{equation}\label{eq:R-Gamma}
R \approx 2\Gamma^2cT_v = 6\times10^{13}\Gamma_2^2T_{v,-1}~\rm{cm}.
\end{equation}
Effect (i) increases the threshold to $\gamma\gamma$-annihilation in terms of the observed photon energy (i.e. decreases $f_{\gamma\gamma}(E)$)
while effect (ii) reduces the required $n_\gamma$. For a power-law photon spectrum $dN/dE\propto E^{-\alpha}$, 
$\tau_{\gamma\gamma}(E)\propto L_0 E^{\alpha-1}/\Gamma^{2\alpha}R\to L_0 E^{\alpha-1}/\Gamma^{2\alpha+2}T_v$ 
\citep[assuming Eq.~(\ref{eq:R-Gamma});][]{Granot+08}, where $L_0= EL_E(E=m_ec^2)$.}

Depending on $\Gamma$, and other intrinsic parameters \citep[e.g.][]{Vianello+17} such as the radiated power $L_\gamma$, $T_v$, 
and $R$ if $R\neq R(T_v)$ \citep[see e.g.][]{GZ08}, the energy where the outflow becomes opaque to $\gamma\gamma$ absorption can be 
pushed to $E\gg E_{\rm pk}$, at which point the non-thermal spectrum is either exponentially suppressed or manifests a smoothly 
broken power-law \citep{Granot+08}.

The existence of a high-energy spectral cutoff occuring due to intrinsic $\gamma\gamma$-opacity has important 
implications. Since the bulk-$\Gamma$ of the outflow is hard to obtain and observations of the highest energy 
photons without a cutoff provide only a lower limit, measuring a spectral cutoff instead yields a direct estimate 
\citep[e.g.][]{FEH93,WL95,BH97,LS01,Razzaque+04,Baring06,MI08,Granot+08,GZ08}. 
So far, a high-energy cut-off has only been observed in a handful 
of sources, e.g. GRB$\;$090926A \citep{Ackermann+11}, and GRBs 100724B \& 160409A \citep[][for a more complete list see \citealt{Tang+15}]{Vianello+17}.
%
Most analytic works  employ a simple one-zone model with an isotropic (comoving) radiation field \citep[e.g.][hereafter LS01]{LS01} and obtain $\Gamma$ from 
the condition that the cutoff energy $E_c$ is given by $E_c=E_1$ where $\tau_{\gamma\gamma}(\Gamma,E_1)\equiv1$.
However, detailed analytic and numerical treatments of the $\gamma\gamma$-opacity near the dissipation region, which account 
for the space, time and direction dependence of the radiation field, by \citet[][hereafter G08]{Granot+08} and \citet{Hascoet+12}, 
respectively, have shown that the actual estimate of $\Gamma$ should be lower by a factor of $\sim2$. 

What was neglected so far in all works is the effect of $e^\pm$-pairs that are produced, and in particular pair cascades, on the scattering opacity and further redistribution of the radiation field energy by Comptonization. Its neglect stems from the inherent non-linearity 
associated with developing pair cascades, which is hard to treat self-consistently using a semi-analytic 
approach and requires a numerical treatment. Highly luminous compact sources with $\tau_{\gamma\gamma}\gg1$ naturally 
develop high Thomson scattering optical depth $\tau_T\gg1$ due to resultant $e^\pm$-pairs that can significantly modify the source spectrum via Comptonization \citep{GFR83}.

Time-dependent numerical models of GRB prompt emission phase \citep[e.g.][]{PW05,VBP11,GT14} self-consistently account for 
$\gamma\gamma$ annihilation, automatically produce the spectral attenuation at comoving energies $E' > m_ec^2$, and account for the enhanced 
scattering opacity due to pair production. This effect will be studied in more detail in a companion paper (Gill \& Granot 2017, in prep.).
Here we use a time-dependent kinetic code to study how $e^\pm$-pairs affect the position of the 
cutoff that arises due to $\gamma\gamma$-opacity. The code includes Compton scattering, cyclo-synchrotron emission 
and self-absorption, pair-production and annihilation, Coulomb interaction, adiabatic cooling, and photon escape. 
In \S~2 we review a simple one-zone model of $\gamma\gamma$ annihilation 
opacity and derive estimates for the scattering optical depth of $e^\pm$-pairs in the optically thick and thin regimes. We construct a general 
model of a magnetized dissipative relativistic outflow in \S~3 in which the prompt emission is attributed to synchrotron 
emission by relativistic electrons and $e^\pm$-pairs. In \S~4 we discuss the implication of our results.


\section{Scaling Relations From a One-Zone Model}
We consider a simple \textit{one-zone} model where
the emission region is uniform with an isotropic radiation field (in its comoving frame). We denote dimensionless photon energies by $x\equiv E/m_ec^2$. The observed prompt emission photon-number spectrum at energies above the 
$\nu F_\nu$-peak, $x_{\rm pk} = E_{\rm pk}/m_ec^2$, can be described by a power-law,
\begin{equation}
\frac{dN}{dA~dT~dx} = N_0\fracb{x}{x_0}^\beta~,\quad\quad x_{\rm pk}<x_0<x<x_{\rm max}
\end{equation}
where $dA \to 4\pi d_L^2(1+z)^{-2}$ for isotropic emission in the CEF,
$dT$ is the differential of the observed time, and $N_0$ 
$[{\rm cm}^{-2}~{\rm s}^{-1}]$ is the normalization. It was shown by LS01 that for 
a given test photon energy, $x_t$, the optical depth due to $\gamma\gamma$-annihilation is
\begin{eqnarray}\label{eq:tauggLS01}
\tau_{\gamma\gamma} &=& (1+z)^{-2(1+\beta)}\hat\tau \Gamma^{2\beta-2}x_t^{-(1+\beta)}\;, 
\\
\hat{\tau} &=& \frac{(11/180)\sigma_Td_L^2N_0x_0^{-\beta}}{c^2\Delta T(-1-\beta)} 
= \frac{(11/180)\sigma_T\tilde L_0 x_0^{-(2+\beta)}}{4\pi c^2\Delta T(-1-\beta)}\;,\quad
\end{eqnarray}
where $\tilde L_0 \equiv L_0(x_0)/m_ec^2 = 4\pi d_L^2 x_0^2N_0$ and $L_0(x_0)$ is the radiated isotropic equivalent luminosity in 
the CEF at $x=x_0$. 
This equation can be used to define the critical photon energy $x_1$ at which $\tau_{\gamma\gamma}(x_1)\equiv ֿ1$. 
If the latter is indeed identified with the observed cutoff energy, $x_c\approx x_1$, this allows us to determine $\Gamma$, 
\begin{equation}
\Gamma_{\rm min}\equiv\Gamma(x_c) = (1+z)^{(-1-\beta)/(1-\beta)}\hat\tau^{1/(2-2\beta)}x_c^{(-1-\beta)/(2-2\beta)}\;.
\label{eq:Gamma1}
\end{equation}
If no spectral cutoff is observed and the power-law extends up to an energy $x_{\rm max}$, then 
Eq.~(\ref{eq:Gamma1}) yields 
a lower limit $\Gamma_{\rm min}=\Gamma(x_c=x_{\rm max})$.

\begin{figure}
\centering
\includegraphics[width=0.32\textwidth]{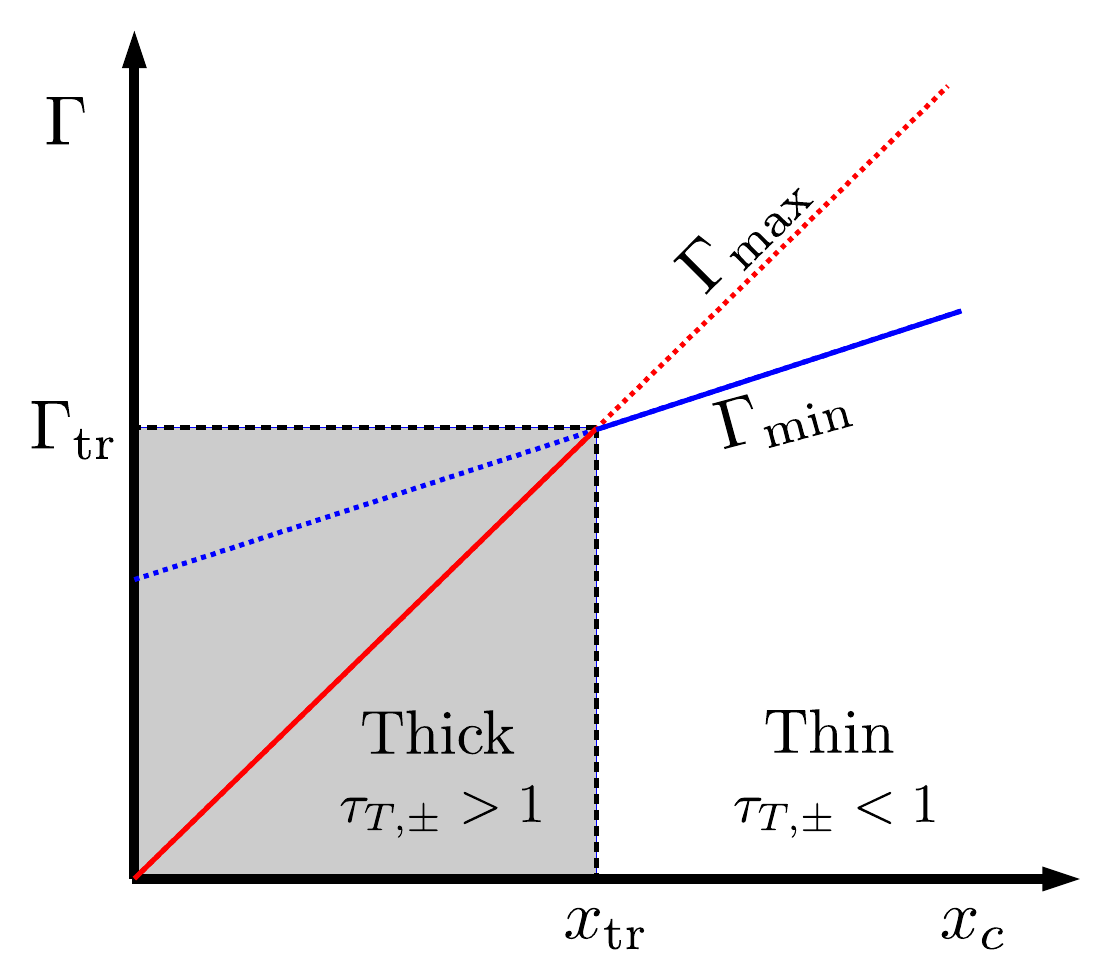}
\caption{Schematic illustration of the two regimes -- Thomson-thick and Thomson-thin -- as defined in the 
text. The bulk Lorentz factor $\Gamma$ of the emitting region is given by $\Gamma=\min(\Gamma_{\rm max},\Gamma_{\rm min})$ and shown by 
solid lines.}
\label{fig:thick-thin}
\end{figure}

In the comoving frame, test photons of energy $x'_t$ have the highest probability to annihilate with 
other photons with energies just above the pair-production threshold, $x'_{\rm an}\approx1/x'_t$, since the cross section decreases well above $x_{\rm an}'$ and vanishes below $x'_{\rm an}$. Therefore, test photons 
of energy $x'_t > x'_{\rm sa}=1$ can self-annihilate whereas photons of energy 
$x'_t<x'_{\rm sa}$ cannot. This has an important consequence for spectra with $\beta < -1$, which is generally 
the case for the prompt-GRB spectrum. In this case, $xdN/dx\propto x^{1+\beta}$ declines with photon
energy $x$, and lower energy photons outnumber higher energy photons. This asymmetry in photon number defines two important regimes (shown in Fig.~\ref{fig:thick-thin}) as follows.

(i) \textit{Thomson-Thick}: In this regime $x_1'<x_{\rm sa}'=1$ so that test photons in the energy range $x_1'<x_t'<x_{\rm sa}'$ initially face $\tau_{\gamma\gamma}(x'_t)>1$, but they cannot self-annihilate. Instead, they can annihilate only with higher energy photons, $x'\geq x'_{\rm an}\approx 1/x'_t$, but since they outnumber these higher energy photons they quickly annihilate almost all of them, which brings down $\tau_{\gamma\gamma}(x'_t)$ below 1. This results in a spectral cutoff at $x'_c = x'_{\rm sa}=1 \Leftrightarrow x_c = \Gamma/(1+z)$, i.e. in this regime $\Gamma=\Gamma_{\rm max}=(1+z)x_c$.
The notation $\Gamma_{\rm max}$ 
was chosen since it is the maximal possible $\Gamma$ for a given cutoff energy $x_c$ due to $\gamma\gamma$-annihilation alone. 

Each annihilating photon-pair produces an $e^\pm$-pair, so the Thomson optical depth of the pairs (ignoring pair-annihilation) in this regime is $\tilde\tau_{T,\pm}= \frac{\sigma_T}{\sigma_{\gamma\gamma}}\tau_{\gamma\gamma}(x'_t=1)$ where $\frac{\sigma_T}{\sigma_{\gamma\gamma}}\approx\frac{180}{11}$. 
Using Eq.~(\ref{eq:tauggLS01}),
\begin{equation}
\tilde\tau_{T,\pm} = \frac{180}{11} \frac{\hat\tau x_c^{\beta-3}}{(1+z)^4} 
= \frac{180}{11}\left[\frac{\Gamma_{\rm min}(x_c)}{\Gamma_{\rm max}(x_c)}\right]^{2(1-\beta)}~,
\end{equation}
where the last equality follows from Eq.~(\ref{eq:Gamma1}). The ratio $\Gamma_{\rm min}(x_c)/\Gamma_{\rm max}(x_c)$ becomes 
unity at the transition energy (LS01),
\begin{equation}
x_{\rm tr} = \left[(1+z)^{-4}\hat{\tau}\right]^{1/(3-\beta)}
\Longleftrightarrow\Gamma_{\rm tr} = \left[(1+z)^{-1-\beta}\hat{\tau}\right]^{1/(3-\beta)}\;,
\end{equation}
which corresponds to $\Gamma_{\rm tr}=(1+z)x_{\rm tr}$. Therefore, for cutoff energies 
$x_c < x_{\rm tr}$, $\Gamma_{\rm min}(x_c) > \Gamma_{\rm max}(x_c)$ and $\tilde\tau_{T,\pm}\gg1$, when $\beta < -1$. In order to 
arrive at this result, the annihilation of $e^\pm$-pairs has been completely ignored, which would certainly 
modify the scattering opacity.

(ii) \textit{Thomson-Thin}: In this regime $1 = x_{\rm sa}'<x_1'$, so photons of energies $x_{\rm sa}'<x'_t<x_1'$ can self-annihilate but have $\tau_{\gamma\gamma}(x'_t)<1$ and only such a small fraction of them indeed annihilate. However, photons of energies $x'_t>x'_1$ face $\tau_{\gamma\gamma}(x'_t)>1$ and almost all of them do annihilate, leading to a cutoff at $x'_c=x'_1$. Hence, in this regime $\Gamma=\Gamma_{\rm min}$ (the minimal possible $\Gamma$ for a given $x_c$). 
Following the discussion above, here $\tilde{\tau}_{T,\pm}\approx\frac{180}{11}\tau_{\gamma\gamma}(1/x'_1)$, or using Eqs.~(\ref{eq:tauggLS01}) \& (\ref{eq:Gamma1}),
\begin{equation}
\tilde\tau_{T,\pm} = \frac{(x'_1)^{2+2\beta}}{11/180} = \frac{180}{11}\left[\frac{(1+z)^4}{\hat{\tau}\,x_c^{\beta-3}}\right]^\frac{1+\beta}{1-\beta}
= \frac{180}{11}\left[\frac{\Gamma_{\rm max}(x_c)}{\Gamma_{\rm min}(x_c)}\right]^{2+2\beta}\;.
\end{equation}
 This Thomson-thin regime corresponds to $x_c > x_{\rm tr}$ and $\Gamma > \Gamma_{\rm tr}$, since here $\Gamma_{\rm min}(x_c) < \Gamma_{\rm max}(x_c)$ which implies $\tilde\tau_{T,\pm}<1$ when $\beta < -1$.

\section{Dissipation in a Relativistic Outflow}
We consider the evolution of a cold, mildly magnetized, \textcolor{black}{expanding spherical shell} 
coasting at a constant $\Gamma=(1-\beta^2)^{-1/2}$ with a constant lab-frame radial width $\Delta$.
At a lab-frame time $t$ the front edge of the ejecta shell is at a radial distance 
$R = \beta ct$ from the central source. \textcolor{black}{Following the relation given in Eq.~(\ref{eq:R-Gamma}), 
the dissipation episode is assumed to start at $R=R_0=6\times10^{13}\Gamma_2^2$~cm} (for brevity, hereafter estimates are given 
for fixed intrinsic parameters: $L_{52}=100$, $T_{v,-1}=1$, magnetization $\sigma=0.1$, 
electron fraction $Y_e=0.5$, but show the explicit dependence on $\Gamma$) and end at 
$R = R_f = R_0+\Delta R$ with $\Delta R = R_0$, i.e. after one dynamical time, but 
the shell can still radiate also at $R>R_f$. 
The rise and decay times of the resulting pulse in the GRB lightcurve are 
$t_{\rm rise}\simeq \Delta R/2c\Gamma^2$ and $t_{\rm decay}\simeq R_f/2c\Gamma^2$, 
so one can in principle use this to determine both $R_0$ and $\Delta R$ if $\Gamma$ can be independently inferred.
The outflow carries a magnetic field of comoving strength $B'\approx4\times10^5\Gamma_2^{-3}\;$G 
and kinetic energy dominated by baryons.

Depending on the efficiency of the dissipation mechanism, a fraction $\varepsilon_{\rm rad}=0.5$ of 
the total power $L_j$ carried by the outflow is converted into radiation, such that the observed isotropic-equivalent luminosity is 
$L = \varepsilon_{\rm rad}L_j=L_{52}10^{52}~{\rm erg~s}^{-1}$. This corresponds to a comoving compactness 
$\ell_0' \approx 2.7\times10^4\Gamma_2^{-5}$ at the dissipation radius $R_0$. The initial Thomson scattering 
optical depth of baryonic electrons is $\tau_{T0} \approx 18\Gamma_2^{-5}$.
A fraction $\eps_{\rm nth}\approx0.87\Gamma_2^{-1}$ of which have 
$\tau_{T,\rm nth}$ and are assumed to be accelerated to a 
power law energy distribution, $n_e'(\gamma_e) \propto \gamma_e^{-q}$ for $\gamma_m < \gamma_e < \gamma_M$, with $\langle\gamma_e\rangle_{\rm nth}$ chosen so that 
the pitch angle averaged synchrotron peak energy of fast cooling electrons yields 
$E_{{\rm p},z} = (1+z)E_{\rm pk}=\Gamma E'_{\rm pk}= 500E_{{\rm p},2.7}\;$keV. The relativistically hot electrons 
are injected with constant power $L$, and then loose all their energy to synchrotron radiation and 
inverse-Compton scattering (ICS) of soft seed photons to high energies.
The remaining fraction $1-\varepsilon_{\rm nth}$ of baryonic electrons with Thomson optical 
depth $\tau_{T,\rm th}$ stay cold $(k_BT'/m_ec^2 \equiv \theta' = 10^{-2})$ 
and form a thermal distribution. The full details of the model will be provided elsewhere (Gill \& Granot 2017, in prep.).
\subsection{Effect of thermal Comptonization \& Pair Annihilation on $x_c$ in the Thomson-thick regime}
\begin{figure}
    \centering
    \includegraphics[width=0.38\textwidth]{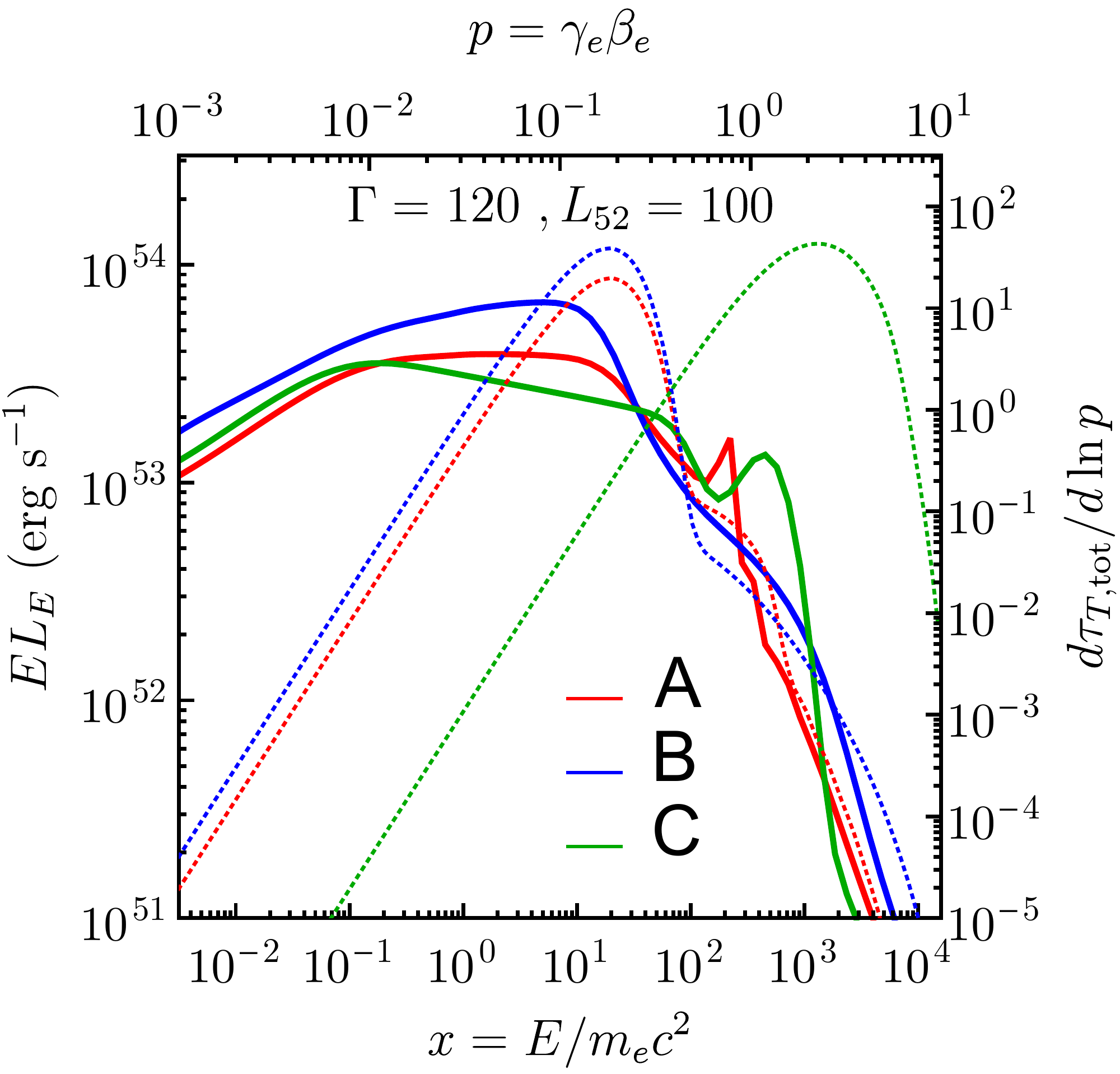}
    \includegraphics[width=0.33\textwidth]{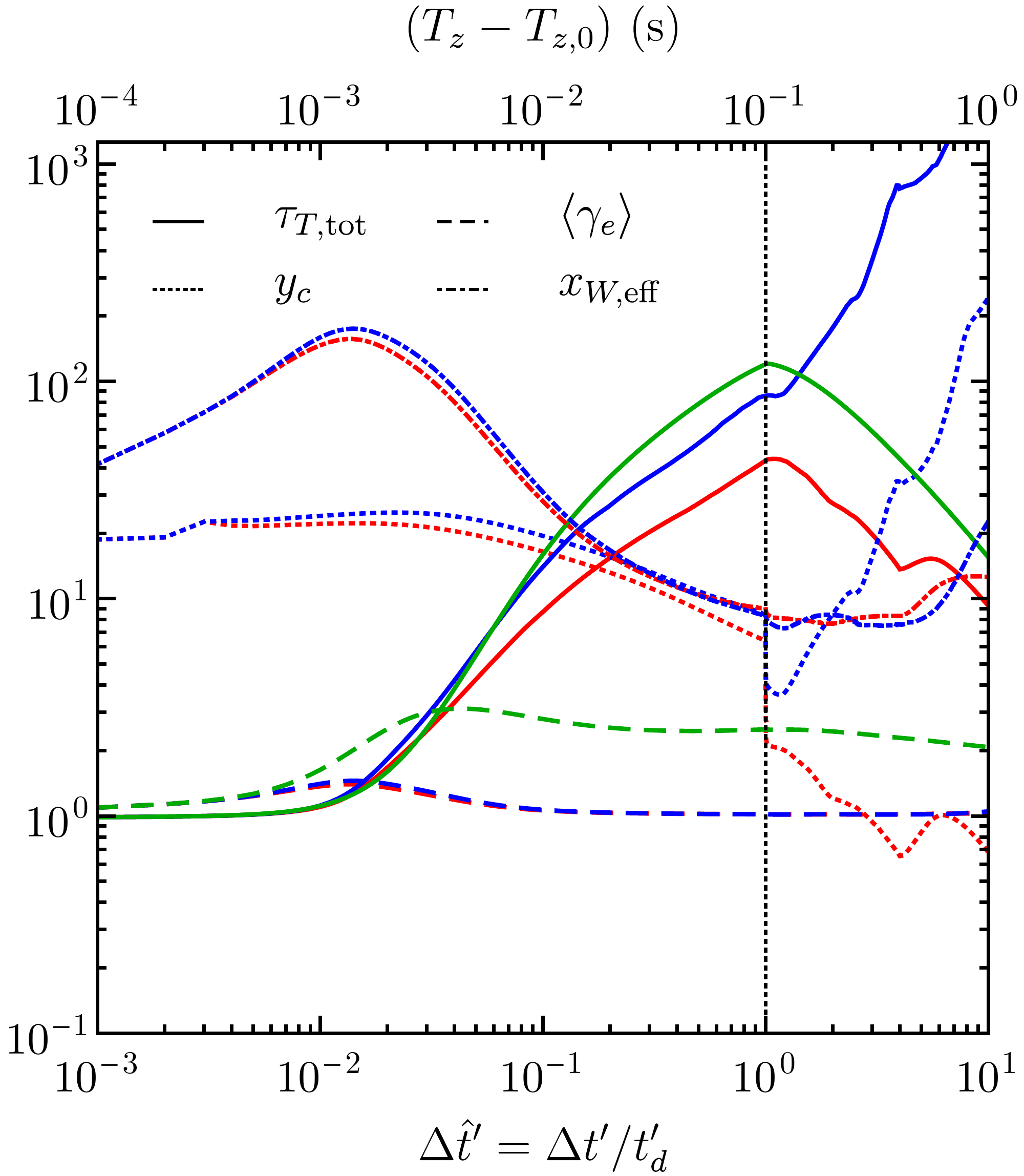}
    \caption{(\textbf{Top}): \textcolor{black}{CEF spectrum from a spherical shell} (solid) and corresponding electron 
    energy distribution (dotted) after one dynamical time, $\Delta t' = t_d'=R_0/\Gamma c\Leftrightarrow\Delta T_z=T_v$, 
    for three cases: {\color{black}(A)} All processes turned on, {\color{black}(B)} No pair-annihilation, {\color{black}(C)} 
    no Compton scattering. (\textbf{Bottom}): Time evolution of some key parameters: the total Thomson depth 
    $\tau_{T,\rm tot}=\tau_{T,\rm th}+\tau_{T,\rm nth}+\tau_{T,\pm}$, Compton $y_C$ parameter, average Lorentz factor 
    of $e^\pm$-pairs $\langle\gamma_e\rangle$, and the effective energy of the Wien peak $x_{W,\rm eff}$.}
    \label{fig:cs-pa-effect}
\end{figure}

In a single ICS event the energy of a soft seed photon ($x_0'$) is amplified by a constant factor, such that 
the scattered photon has energy $x'=(1+A)x_0'$. For ultra-relativistic electrons with $\gamma_e\gg1$, the mean fractional change in the seed photon's energy is $A=\frac{4}{3} \gamma_e^2$ in the Thomson-limit ($x'_0\gamma_e<1$). If the scattering electrons have a Maxwellian distribution with temperature $\theta'$, then thermal Comptonization yields (when neglecting downscattering) 
$A=16\theta'^2+4\theta'$ \citep[e.g.][]{RL79,PSS83}, which is valid for both non-relativistic ($\theta'<1$) and relativistic ($\theta'>1$) electrons. The importance of 
multiple ICSs in modifying the seed spectrum is gauged by the magnitude of the Compton parameter,\footnote{Usually $\max(\tau_T,\tau_T^2)$ is taken for the mean number of scatterings $N_{\rm sc}$ instead of $\tau_T$, but here 
$\tau_T\propto R^{-2}$ due to the shell's expansion so that $N_{\rm sc}\sim\tau_{T0}$ as it is dominated by 
the first dynamical (or radius-doubling) time.} $y_C=A\tau_T$, 
where $\tau_T$ is the electron Thomson optical depth. After multiple scatterings, upon its escape, 
the seed photon's energy is amplified to $x'_f\sim x_0'e^{y_C}$ (for $x'_f\ll4\theta'$). Thus, when 
$y_C>1$ and $\tau_T>1$ Comptonization becomes important, and for $y_C\gg1$ it ``saturates'' and forms 
a Wien peak at $x'_W=3\theta'$.

Fig.~\ref{fig:cs-pa-effect} shows the central-engine frame (CEF) spectrum at the end of one dynamical time (top-panel), stressing the spectral changes brought by Comptonization and pair-annihilation. The corresponding 
electron energy distribution is predominantly thermal in all three cases due to the high total 
$\tau_{T,\rm tot}=\tau_{T,\rm th}+\tau_{T,\rm nth}+\tau_{T,\pm}$ (lower-panel). Initially $\tau_{T,\rm tot}=\tau_{T,\rm th}=(1-\varepsilon_{\rm nth})\tau_{T0}\approx1$ 
which builds up over the dynamical time, $t_d'=R_0/\Gamma c$, due to injection of relativistic electrons 
and subsequent production of $e^\pm$-pairs that dominate $\tau_{T,\rm tot}$ when $\ell'\gg1$. It suffers a sharp decline at $\Delta t'=t_d'$ after which 
injection of electrons ceases and the hot pairs cool and annihilate with the thermal pairs. This is not 
so when pair-annihilation is turned off. In all cases (except with no ICS), $y_C\gg1$ which 
results in saturated Comptonization. The position of the Wien peak in the observer frame is obtained 
from $x_W\simeq2\Gamma x_W'=6\Gamma\theta'$ (the factor of 2 results from higher weight given to on-axis ($\theta_\Gamma=0$)
emission upon integration over the equal arrival time surface \citep{GPS99} since $L/L'=\delta_D^4$ where 
$\delta_D\approx2\Gamma/[1+(\Gamma\theta_\Gamma)^2)]$ is the Doppler factor and $\theta_\Gamma$ is the 
angle measured from the line of sight). 

The temperature of the thermal pairs is related to their mean energy, 
$\langle\gamma_e\rangle=[3\theta' K_2(1/\theta')+K_1(1/\theta')]/[2\theta' K_1(1/\theta')+K_0(1/\theta')]$ 
\citep{PSS83}, where $K_n$ are the modified Bessel functions of the second-kind. Since at early times the 
particle distribution is quasi-thermal that transforms into predominantly thermal over time, the 
$\langle\gamma_e\rangle-\theta'$ relation only yields the ``effective'' temperature of the $e^\pm$-pairs, 
and consequently an effective energy for the Wien peak ($x_{W,\rm eff}$). At the end of the dynamical 
time, when $\tau_T\gg1$, this approximation becomes more exact as the particles and photons come into 
thermal equilibrium. When Compton cooling of the injected relativistic electrons and mildly relativistic 
$e^\pm$-pairs is switched off, the hot particles share their energy with the much cooler thermal (baryonic) 
electron distribution via Coulomb interactions. This has the effect of heating up the thermal distribution 
which yields higher particle temperatures by the end of the dynamical time and broadens the pair annihilation 
line. In contrast, ICS of hot electrons on soft synchrotron photons helps regulate the temperature of the 
particle distribution to much lower ($\theta'<1$) values, which also yields a much sharper annihilation feature.

How far below unity can the temperature of a pair-dominated plasma drop? Many works have tried to 
understand thermal pair equilibria of mildly relativistic \citep{Svensson84} and relativistic 
plasmas \citep[][]{Lightman82,Svensson82}. 
By solving the pair balance equation, where pair production balances pair annihilation in steady-state, it 
was realized that no equilibrium exists for $\theta'>\theta'_{\rm max}=24$ when $\ell'\ll1$ and for 
$\theta'\gtrsim0.4$ when $\ell'>\ell_{\rm WE}(\theta')\gg1$ \citep{Svensson84}. At $\ell'\gg1$, when 
Comptonization dominates over photon emission/absorption and escape, the pairs establish a Wien equilibrium 
where the compactness of the radiation field depends uniquely on the pair temperature for 
$\theta'\lesssim0.4$, such that $\ell_{\rm WE}'(\theta') = 4\sqrt{2\pi}\theta'^{5/2}\exp(1/\theta')$ 
\citep{Svensson84}. Hence, the temperature of the non-relativistic thermal pairs decreases below unity 
logarithmically with $\ell'$. 
This trend continues until a local thermodynamic equilibrium is established due to true photon 
emission/absorption processes (e.g. cyclo-synchrotron emission and self-absorption).

From the condition of pair equilibrium, the relation between $\ell'$ and $\tau_{T,\pm}$ can be obtained 
when $\ell'\gg1$. Under the assumption that all of the injected energy at this stage goes into producing 
hard photons ($x'>1$) that can annihilate with other soft ($x'<1$) photons as well as self-annihilate, 
the rate of pair production is $\dot n_+'=L/(4\pi R^3\Gamma m_ec^2)=c\Gamma^2\ell'/\sigma_TR^2$. These pairs 
then annihilate the cooler thermal pairs at the rate $\dot n_A'\sim \sigma_Tcn_+'^2$. In equilibrium 
$\dot n_+'=\dot n_A'$, which yields up to a factor of order unity $\tau_{T,\pm,\rm Eq}\sim\sqrt{\ell'_0}(R_0/R)$ 
\citep[e.g.][]{PW04}, where the density dilution due to expansion is reflected by the ratio of radii. 
In the top-panel of Fig.~\ref{fig:Gamma-LS01-compare} we show $\tau_{T,\pm,\rm eq}$ along with the total 
Thomson depth of particles, which is dominated by that of pairs, at the end of a dynamical time from simulations 
without pair annihilation and Compton scattering. We find that $\tau_{T,\rm tot}\approx\tau_{T,\pm}\sim\tau_{T,\pm,\rm eq}$ for 
$\ell'\gg 1$ when all radiative processes are included. The agreement is approximate since the pair-photon plasma 
hasn't established a steady state. When pair-annihilation is switched off, we find 
$\tilde\tau_{T,\rm tot}\approx2\tau_{T,\rm tot}$ after one dynamical time over a wide range of compactness.

\subsection{Comparison With One-Zone Analytic Model Predictions}

\begin{figure}
    \centering
    \includegraphics[width=0.38\textwidth]{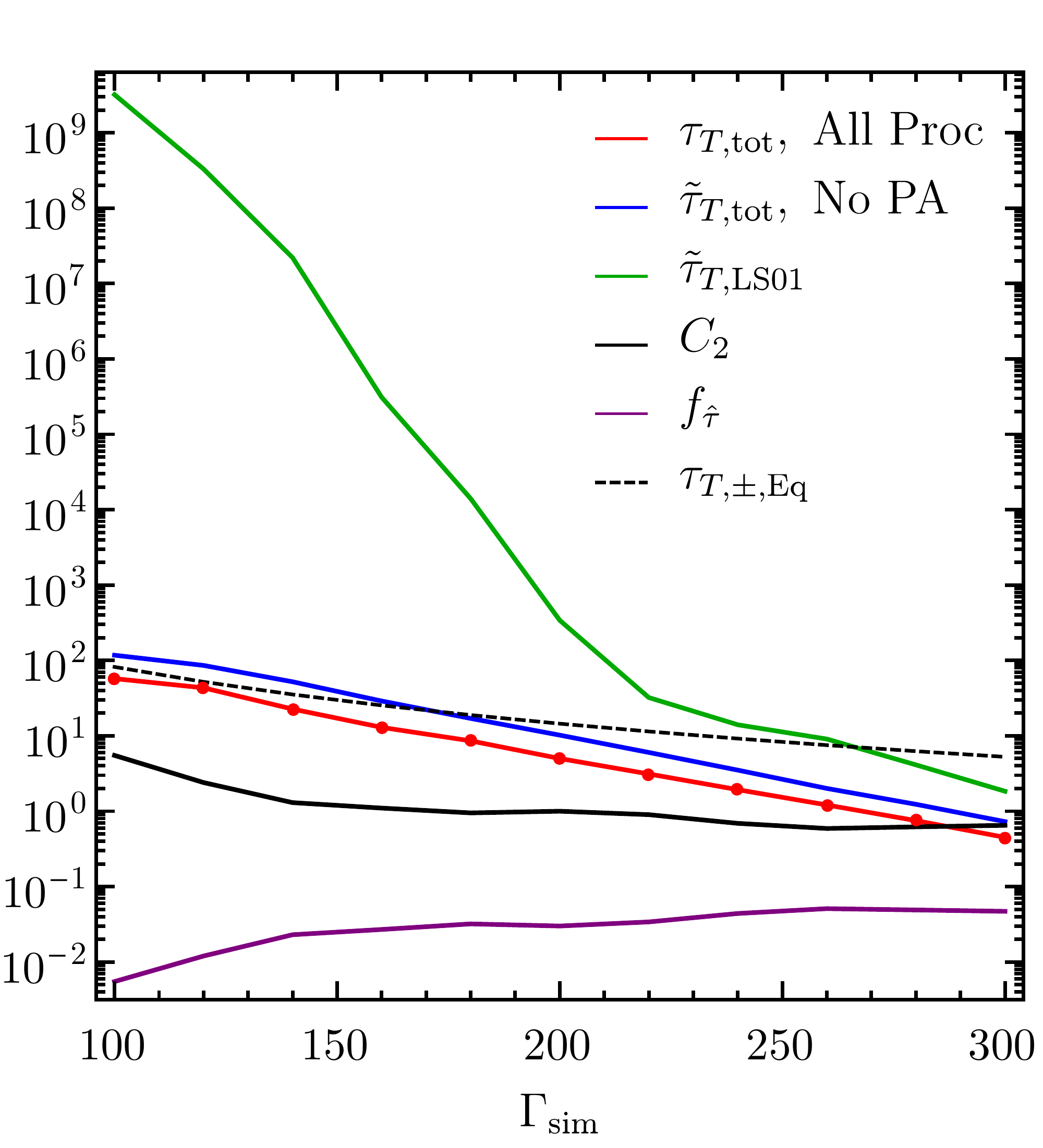}
    \includegraphics[width=0.4\textwidth]{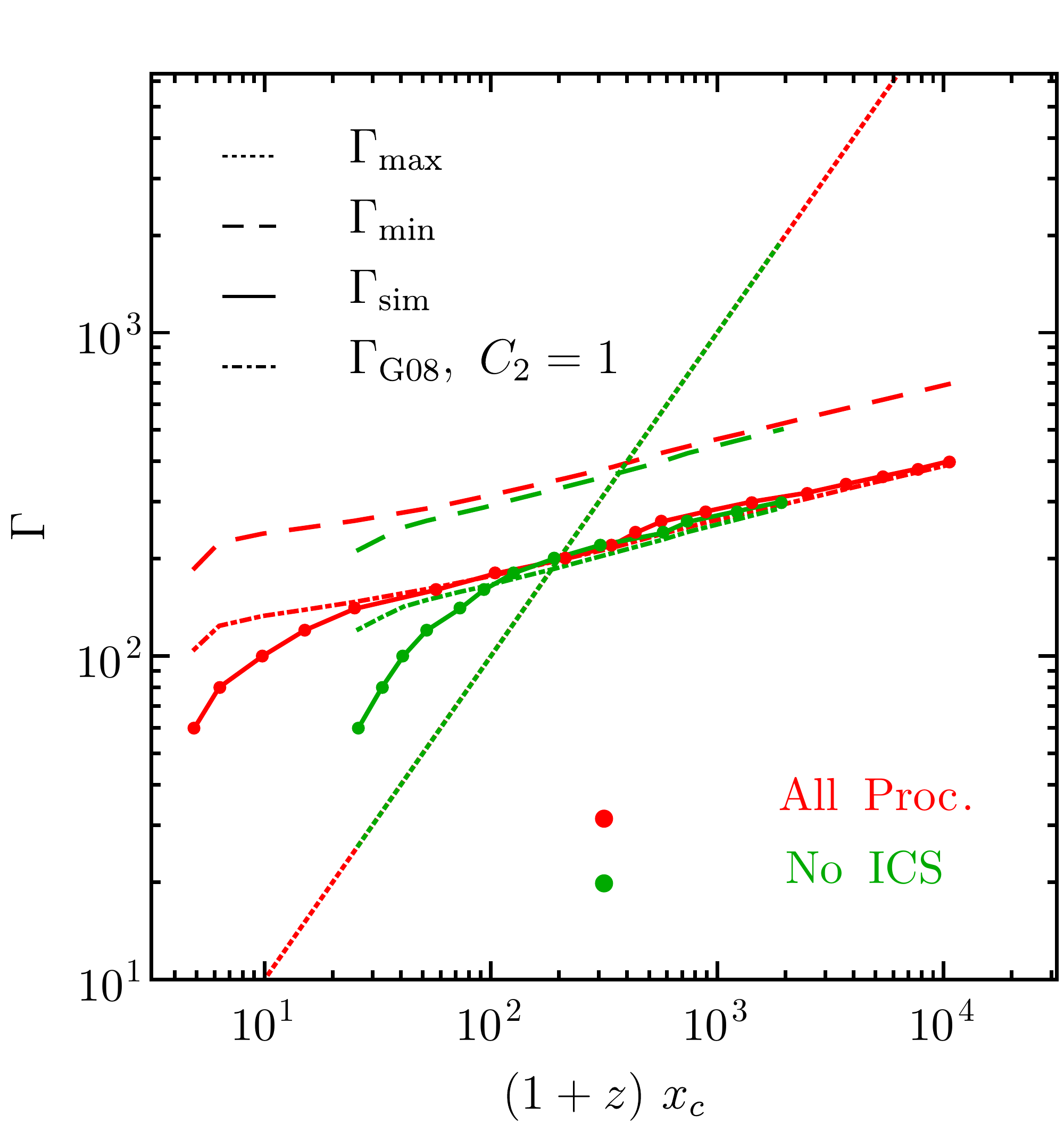}
    \caption{(\textbf{Top}) Comparison of the Thomson depth from simulations with all processes active and with 
    no pair annihilation (PA) to that obtained from the analytic model of LS01. See text for definition of all 
    quantities. (\textbf{Bottom}) Comparison of $\Gamma$ from the simulation $\Gamma_{\rm sim}(x_c)$ 
    to the prediction of the LS01 analytical model $\Gamma_{\rm min}(x_c)$ and $\Gamma_{\rm max}=(1+z)x_c$ 
    (compare with Fig.~\ref{fig:thick-thin}). 
    $\Gamma$ predicted by Eq.~(126) of G08 ($\Gamma_{\rm G08}$) for $C_2=1$ is also shown.}
    \label{fig:Gamma-LS01-compare}
\end{figure}

Earlier we outlined two regimes of the one-zone analytic model of LS01 which did not account for annihilation of 
$e^\pm$-pairs. Here we compare the results of our simulations to the predictions of 
LS01. First we need to determine the position of the high-energy cutoff, which along with other spectral parameters 
such as the high-energy spectral slope and normalization, yields an estimate of $\Gamma$. We obtain the position 
of the cutoff energy in the CEF by 
fitting the spectrum to
a Band-function \citep[][]{Band+93} with a 
broken power-law high-energy cutoff \citep[see Eq.~(E5) of][]{Vianello+17}.

In Fig~\ref{fig:Gamma-LS01-compare} we compare the results obtained from the simulations to the 
predictions of the LS01 analytic model and Eq.~(126) of G08. We find that the LS01 model grossly over-predicts 
the Thomson scattering depth of pairs in the Thomson-thick regime. In the Thomson-thin regime, $\tilde\tau_{T,\rm LS01}$ 
asymptotically approaches $\tilde\tau_{T,\rm tot}$ in Fig.~\ref{fig:Gamma-LS01-compare} since the analytical 
model does not account for pair annihilation. In addition, the LS01 model also finds $\Gamma_{\rm min}$ to 
be a factor $\sim2$ larger than the 
simulated value $\Gamma_{\rm sim}$. This result is consistent with the work of \citet{Vianello+17}, where they compare $\Gamma$ 
obtained from the models of G08 and \citet{GT14} that self-consistently produce high-energy spectral breaks to that predicted by 
the LS01 model for GRBs 100724B \& 160509A. The predictions of LS01 can be reconciled with the simulation results by renormalizing 
$\hat\tau\to f_{\hat\tau}\hat\tau$ in Eq.~(\ref{eq:Gamma1}), where $f_{\hat\tau}\equiv\hat\tau_{\rm sim}/\hat\tau_{\rm LS01}$. This ratio is 
shown in Fig.~\ref{fig:Gamma-LS01-compare} where $f_{\hat\tau}\sim0.05$ for $\Gamma>100$.

The main effect of Compton scattering (see Fig.~\ref{fig:Gamma-LS01-compare}) is that $x_c$ 
becomes lower due to downscattering of energetic photons with $x'>4\theta'$ by cold thermal $e^\pm$ pairs. 
We find good agreement between $\Gamma_{\rm sim}$ and $\Gamma_{\rm G08}$ in the Thomson-thin regime for 
$C_2 = 1$, where $C_2$ is an order unity parameter in Eq.~(126) of G08 whose exact value is determined 
numerically. When ICS is switched on, we find $C_2\sim0.5-1$ in order for $\Gamma_{\rm G08}=\Gamma_{\rm sim}$, 
which is consistent with the results of \citet{Vianello+17}; $\Gamma_{\rm G08}$ deviates only slightly (factor of $\sim$2) deep in the 
Thomson-thick regime and remains a reliable estimator of the true $\Gamma$ for $\Gamma_{\rm max}\gtrsim0.1\Gamma_{\rm tr,G08}\Leftrightarrow x_c\gtrsim0.1x_{\rm tr,G08}$, where 
$\Gamma_{\rm tr,G08}=(1+z)x_{\rm tr,G08}$ and $x_{\rm tr,G08}$  are defined as the point where $\Gamma_{\rm G08}=\Gamma_{\rm max}$.

Most importantly, the break in $\Gamma(x_c)$ at $x_c=x_{\rm tr}$ predicted by the LS01 model is shifted to 
lower energies both when ICS is on and off, where ICS shifts the break to even lower energies. 
Also, the slope in the Thomson-thick regime is different in the two cases. In that 
regime, if the photo-pair-plasma has achieved steady-state and manifests a Wien-peak then the $\ell_{\rm WE}'-\theta'$ relation 
can be used to determine $\Gamma(x_c)$. Since neither of the two conditions are fulfilled here this relation is invalid. 
In the absence of ICS, when no downscattering of photons occurs, we find that $\Gamma_{\rm sim}(x_c)$ asymptotically approaches $2\Gamma_{\rm max}(x_c)$ in the Thomson-thick regime when $\ell'\gg1$.

\section{Implications And Discussion}
In many works \citep[e.g.][]{Tang+15} that find the spectral cutoff to lie in the Thomson-thick regime, 
$\Gamma$ is estimated using $\Gamma_{\rm max}$. It is clear from Fig.~\ref{fig:Gamma-LS01-compare} that 
this approach can lead to erroneous results and can underestimate $\Gamma$ by as much as an order of 
magnitude when $\ell'\gg1$. \textcolor{black}{This result is quite general such that it doesn't depend on 
the details of any particular GRB model, 
but only on the compactness of the dissipation region that is set by a combination of three intrinsic 
parameters: $L$, $\Gamma$, $R$. Further 
discussion of this effect in the context of many popular models of GRB prompt emission, e.g. internal shocks, 
magnetic dissipation, and photospheric, will be presented in a companion work (Gill \& Granot, in prep.).}

In the Thomson-thin regime, the simple analytic model over-predicts $\Gamma$ by a factor of $\sim2$
(G08; \citealt{Hascoet+12}). This work shows that the effect of pair cascades on the high-energy spectral cutoff cannot be ignored and, more 
importantly, a model employing the time-dependent evolution of the spectrum must be used to obtain an accurate estimate of $\Gamma$. The 
simple one-zone analytical models lack the requisite complexity to accurately predict $\Gamma$. 

In this work, the cutoff energy is determined for a single pulse after integrating the spectrum over the equal arrival time surface. 
Generally, due to poor photon statistics, observations use several overlapping pulses emerging from different parts 
of the outflow with an order unity spread in $\Gamma$. This introduces some smearing of the cutoff energy and 
sharp annihilation line within a single pulse as well as over several adjoining pulses. This effect will be explored 
in detail elsewhere (Gill \& Granot 2017, in prep.).

The simple one-zone analytic models of e.g. LS01, \citet{Abdo+09b} disagree with the more detailed analytic work of G08 and the results 
presented here due to the following main reasons: (i) They only use the power-law component of the Band function rather than the smoothly 
broken power-law at $x<x_{\rm pk}$ (however see for e.g. \citet{GZ08}; although G08 also uses an infinite power-law but see (ii)). 
For typical spectral indices $\alpha\sim-1$ and $\beta\sim-2$ below and above $x_{\rm pk}$ respectively, 
the number of photons in an infinite power-law are larger by a factor of $x_{\rm pk}/[x\log (x/x_{\rm pk})]$ 
for $x<x_{\rm pk}$ as compared to the Band function. This decrement in photon number reduces $\tau_{\gamma\gamma}$ 
seen by hard photons with $x>1/x_{\rm pk}$. Consequently, the estimated $\Gamma$ is lower. (ii) The assumption of 
(comoving) isotropy of the radiation field in such models yields higher estimates of $\Gamma$. The effect of an 
anisotropic radiation field is to increase the threshold for pair production and decrease the rate of interaction 
due to the typical angle of interaction between photons $\theta_{12}\sim1/\Gamma$. This effect is included in 
G08 which generally finds a lower $\Gamma$. (iii) All analytic models neglect the effect of pair cascades, which 
becomes very important in the Thomson-thick regime.
\section*{Acknowledgements}
RG and JG acknowledge support from the Israeli Science Foundation under Grant No. 719/14. RG is supported by an Open University of Israel Research 
Fellowship. 



\bsp	
\label{lastpage}
\end{document}